\documentclass[a4paper,10pt,twocolumn]{article}
\usepackage[a4paper,right=1.5cm,left=1.5cm,top=2cm,bottom=2.5cm,headsep=0cm,footskip=0.5cm]{geometry}
\usepackage[english,activeacute]{babel}

\usepackage[utf8]{inputenc}
\usepackage{enumitem}
\usepackage{amsmath}
\usepackage{amssymb}
\usepackage{times}
\usepackage{graphicx}
\usepackage{float}
\usepackage{csquotes} 
\usepackage{multirow}
\usepackage{xcolor}
\usepackage{colortbl}
\usepackage[colorlinks=true,allcolors=blue]{hyperref}
\usepackage{titlesec}
\usepackage{subcaption}
\usepackage{float}
\usepackage[font=small,labelfont=bf,justification=justified,singlelinecheck=false]{caption}
\usepackage{adjustbox}
\usepackage{comment}
\usepackage{marvosym}

\titleformat{\section}{\large\bfseries}{\thesection}{1em}{} 
\titleformat{\subsection}{\small\bfseries}{\thesubsection}{1em}{} 
\titleformat{\subsubsection}{\small\bfseries}{\thesubsubsection}{1em}{} 
\titlespacing{\section}{0pt}{\parskip}{\parskip}
\titlespacing{\subsection}{0pt}{\parskip}{\parskip}

\renewcommand\thesection{\arabic{section}.}
\renewcommand\thesubsection{\arabic{section}.\arabic{subsection}.}

\title{\LARGE \textbf{Masked Autoencoder Joint Learning for Robust Spitzoid Tumor Classification}}

\author{\large Ilán Carretero $^{1}$, Roshni Mahtani $^{1}$, Silvia Perez-Deben $^{2}$, José Francisco González-Muñoz $^{2}$,\\ Carlos Monteagudo $^{2}$, Valery Naranjo $^{1,3}$, Rocío del Amor $^{1,3}$
\\
\\
\small $^1$ HUMAN-tech, Universitat Politècnica de València (UPV), Valencia, Spain   \\
\small $^2$ INCLIVA, Universitat de València (UV), Valencia, Spain \\
\small $^3$ Artikode Intelligence S.L., Valencia, Spain  \\
\small \Letter \{ilcarjuc, rmahvas, vnaranjo, madeam2\}@upv.es \\
}
\date{} 

\begin{document}	

\maketitle
\thispagestyle{empty}

\begin{center}
\large
\textbf{Abstract}
\end{center}
\small
\textit{Accurate diagnosis of spitzoid tumors (ST) is critical to ensure a favorable prognosis and to avoid both under- and over-treatment. Epigenetic data, particularly DNA methylation, provide a valuable source of information for this task. However, prior studies assume complete data, an unrealistic setting as methylation profiles frequently contain missing entries due to limited coverage and experimental artifacts. Our work challenges these favorable scenarios and introduces ReMAC, an extension of ReMasker designed to tackle classification tasks on high-dimensional data under complete and incomplete regimes. Evaluation on real clinical data demonstrates that ReMAC achieves strong and robust performance compared to competing classification methods in the stratification of ST. Code is available: \href{https://github.com/roshni-mahtani/ReMAC}{https://github.com/roshni-mahtani/ReMAC}.} 

\normalsize

\section{Introduction}

Spitzoid tumors (ST) are melanocytic neoplasms defined by large spindle and epithelioid cell morphology, coupled with unpredictable clinical behavior \cite{barnhill2006spitzoid}. ST are typically classified into three categories: the benign form, Spitz Nevus (SN); an intermediate entity with uncertain malignant potential, Atypical Spitz Tumor (AST) or Spitz melanocytoma; and the malignant form, Spitz Melanoma (SM) \cite{barnhill2006spitzoid}. Accurate diagnosis is essential, as misclassification may result in severe clinical consequences and inappropriate treatment \cite{orchard1997spitz}.

Epigenetic profiling, particularly through Reduced Representation Bisulfite Sequencing (RRBS), has become a widely adopted next-generation sequencing technique that enriches CpG-dense regions to generate genome-wide methylation maps at single-base resolution \cite{beck2022genome}. Building on this foundation, several studies have leveraged DNA methylation (DNAm) to advance in the stratification of spitzoid tumors \cite{delamor2021deep, gonzalez2023diagnostic}. However, most approaches have generally overlooked the issue of missing values, a recurrent challenge in methylation data stemming from limited sequencing depth or technical variability \cite{seiler2021characterizing}. This limitation highlights the need for computational frameworks that remain accurate and robust when operating on incomplete epigenetic data.

Several state-of-the-art imputers have been developed to address missing values, including discriminative approaches such as MICE \cite{van2011mice} and MIRACLE \cite{kyono2021miracle}, as well as generative models such as GAIN \cite{yoon2018gain} and HI-VAE \cite{hivae}. More recently, ReMasker  \cite{remasker} introduced a self-supervised masked autoencoding (MAE) framework for tabular data imputation, demonstrating strong performance and the ability to learn missingness-invariant representations. Nevertheless, this model has not been tested on high-dimensional data such as epigenetics, nor has it been extended to produce latent representations tailored for classification tasks.

In this work, we propose ReMAC, a framework that extends the \textbf{Re}Masker model by incorporating a \textbf{M}ean \textbf{A}ttribution \textbf{C}lassification (MAC) branch. As a result, ReMAC emerges as a state-of-the-art method capable of producing robust discriminative representations on DNA methylation data to classify spitzoid tumors.

\section{Methodology}

\begin{figure*}[hbt]
\centering
        \includegraphics[width=\textwidth]{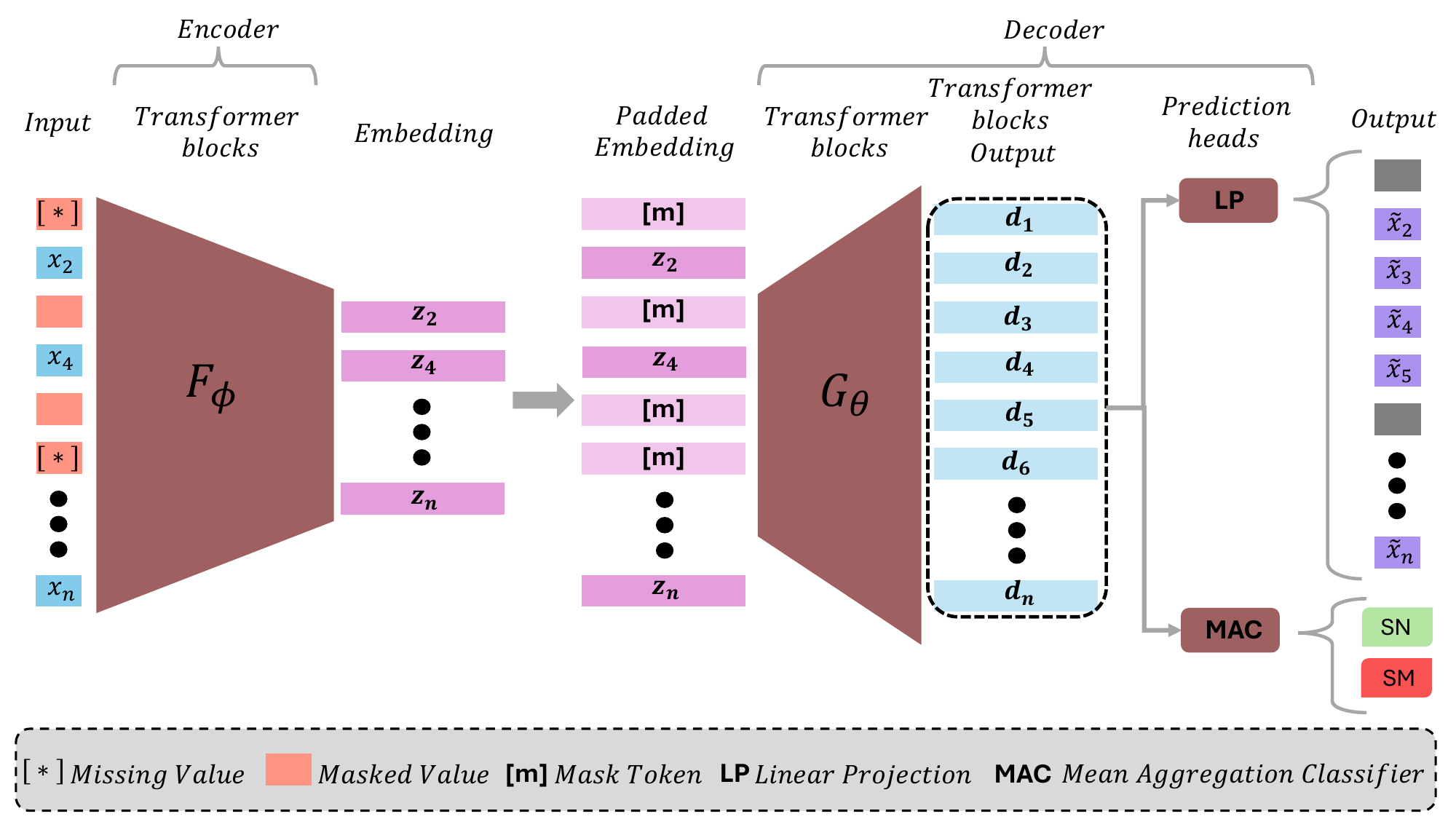}
    \caption{\textit{Method Overview. In this article, we introduce ReMAC, a framework that extends the ReMasker approach by incorporating an aggregation and classification head to learn missingness-invariant discriminative representations for the stratification of spitzoid tumors into Spitz nevus and Spitz melanoma.}}
    \label{fig:framework}
\end{figure*}

An overview of the proposed method is illustrated in Fig. \ref{fig:framework}. The problem formulation and the different components implemented are described in the following subsections. 

\subsection{Problem formulation}

Let $\mathcal{S} = \{(x^{(j)}, y^{(j)})\}_{j=1}^M$ denote the set of tumor samples ($M$), where each sample $x^{(j)} = (x^{(j)}_1, x^{(j)}_2, \dots, x^{(j)}_n) \in \mathbb{R}^n$ is the epigenetic profile represented by $n$ features, and $y^{(j)} \in \{0,1\}$ is the corresponding label. The objective is to learn a mapping $h_\psi : \mathbb{R}^n \rightarrow \{0,1\}$, where label $0$ corresponds to SN and label $1$ corresponds to SM.

\subsection{Masked Autoencoding for tabular data}

ReMasker \cite{remasker} is a masked autoencoding framework for tabular data. An encoder $F_\phi$, implemented as a stack of Transformer blocks, maps the input into a set of latent embeddings $Z = \{z_i\}_{i=1}^n$. 
To model missing patterns, positions selected for masking are added in $Z$ by a learned mask token $\mathbf{m}$. This design enables the encoder to capture contextual feature dependencies through multi-head self-attention. \\

A decoder $G_\theta$, also composed of Transformer blocks, processes the augmented set of embeddings $Z$, and a subsequent linear projection (LP) maps the decoder embeddings $d_i$ to the feature space, producing reconstructions $\tilde{x}_i = \mathrm{LP}(d_i)$. 
The training objective minimizes the mean squared error (MSE) restricted to the union of observed indices $\Omega_o$ and masked-but-originally-observed indices $\Omega_m$:  
\[
\mathcal{L}_{\text{REC}} = \frac{1}{|\Omega_o \cup \Omega_m|} \sum_{i \in \Omega_o \cup \Omega_m} \big(x_i - \tilde{x}_i\big)^2.
\]  
This formulation forces the model to reconstruct true feature values from contextual information, yielding robust latent representations invariant to missingness patterns.

\subsection{Extending ReMasker for classification}

ReMasker learns latent representations invariant to missingness patterns; however, its original design is limited to the imputation task. To enable classification, we introduce a Mean Aggregation Classification (MAC) head. Let $D=\{d_i\}_{i=1}^n$ denote the sequence of decoder embeddings, i.e., the token representations produced by the decoder $G_\theta$. These embeddings are aggregated via mean pooling,
$\;\bar d=\frac{1}{n}\sum_{i=1}^n d_i,$
and a dense layer maps the global representation to class probabilities,
$\;\tilde y=\mathrm{softmax}(W\bar d+b),\;$ with $y\in\{0, 1\}.$
The classification loss is defined as the binary cross-entropy $\mathcal{L}_{\mathrm{CLF}}=\mathrm{BCE}(y,\tilde y)$, and the overall training objective combines reconstruction and classification:
\[
\mathcal{L}_{\mathrm{ReMAC}}=\mathcal{L}_{\mathrm{REC}}+\mathcal{L}_{\mathrm{CLF}}.
\]

This joint objective ensures that the learned representations remain robust to missingness while being explicitly shaped for discriminative classification.

\section{Experimental setting}

\subsection{Dataset}

For this study, a total of 21 formalin–fixed paraffin–embedded (FFPE) tumor samples from patients with spitzoid lesions were analyzed. These included 12 cases classified as Spitz Nevus (SN) and 9 cases classified as Spitz Melanoma (SM). Tumor specimens were collected at the time of surgery and reported to the Department of Anatomic Pathology of the Hospital Clínico Universitario, Valencia (Spain) between 1990 and 2018. The protocol was approved by the Ethical and Scientific Committees of the Hospital Clínico Universitario, and written informed consent was obtained from all patients.

\subsection{Bioinformatic preprocessing}

Sequencing quality was assessed with \emph{FastQC}, and adapters were removed using \emph{Trim Galore!}. Reads were aligned to the human reference genome (hg19) with \emph{Bismark}, followed by methylation calling. Additional quality control metrics were summarized with \emph{MultiQC}, and coverage normalization was carried out using the \emph{methylKit} R package. The resulting methylation values ranged from 0 to 1 with no missing values.

\subsection{Implementation and validation protocol}

All experiments were conducted under a 4-fold stratified cross-validation regime. The proposed method, as well as competing state-of-the-art approaches, were implemented in Python 3.10 using standard libraries for data processing and modeling (e.g., \emph{NumPy} 1.22.2, \emph{pandas} 1.5.3, \emph{scikit-learn} 1.2.0). To ensure reproducibility of both code and results, pseudo-random seeds were fixed and experiments were executed within Docker, specifically using the \emph{PyTorch} 23.10 container image.

\section{Results}

\subsection{Results on the complete dataset}

The results of different classification models on the complete dataset, together with ReMAC, are reported in Table \ref{tab:results_complete_data}. The proposed method is compared with conventional machine learning approaches, including K-Nearest Neighbors (KNN), Logistic Regression (LR), Support Vector Machine (SVM), Random Forest (RF), and eXtreme Gradient Boosting (XGB), as well as with deep learning models such as a Multilayer Perceptron (MLP), an Autoencoder with a classification head (AE+MLP), and TabNet \cite{arik2021tabnet}. ReMAC achieves strong performance in the classification of spitzoid tumors under the complete data regime, outperforming the competing models in two out of the four reported metrics.

\begin{table}[ht]
\centering
\begin{adjustbox}{width=0.47\textwidth}
\begin{tabular}{lcccc}
\hline
\multicolumn{1}{c}{MODEL} & ACC                  & SEN                  & SPE                  & AUC                  \\ \hline
KNN                       & 0.73 ± 0.15          & 0.63 ± 0.48          & 0.83 ± 0.19          & 0.77 ± 0.13          \\
LR                        & 0.73 ± 0.22          & 0.63 ± 0.48          & 0.83 ± 0.19          & 0.82 ± 0.14          \\
SVM                       & 0.68 ± 0.15          & 0.63 ± 0.48          & 0.75 ± 0.17          & \textbf{0.86 ± 0.10} \\
RF                        & 0.68 ± 0.15          & 0.63 ± 0.48          & 0.75 ± 0.17          & 0.85 ± 0.11          \\
XGB                       & 0.68 ± 0.15          & 0.50 ± 0.41          & 0.83 ± 0.19          & 0.58 ± 0.10          \\
MLP                       & 0.81 ± 0.16          & 0.79 ± 0.25          & 0.83 ± 0.19          & \textbf{0.86 ± 0.10} \\
AE+MLP                    & 0.81 ± 0.02          & \textbf{0.92 ± 0.17} & 0.75 ± 0.17          & \textbf{0.86 ± 0.10} \\
TabNet                    & 0.72 ± 0.22          & 0.71 ± 0.30          & 0.75 ± 0.43          & 0.73 ± 0.18          \\
\rowcolor[HTML]{EFEFEF}
ReMAC (\textit{Ours})                     & \textbf{0.86 ± 0.10} & 0.67 ± 0.24          & \textbf{1.00 ± 0.00} & \textbf{0.86 ± 0.10} \\ \hline
\end{tabular}
\end{adjustbox}
\caption{\textit{Comparison of machine learning and deep learning models on the complete dataset (no missing values). Reported metrics include accuracy (ACC), sensitivity (SEN), specificity (SPE), and area under the curve (AUC). The best value for each metric is highlighted in bold, and the proposed method is shown with a gray background.}}
 \label{tab:results_complete_data}
\end{table}

\subsection{Results on incomplete datasets}

Table \ref{tab:results_incomplete_data} reports the performance of classification models robust to missing data, together with ReMAC, under pseudo-randomly generated missing-value (\%MV) regimes. We considered widely used models for classification with incomplete data, namely Histogram-based Gradient Boosting (HistGB), XGB, and CatBoost. Across most metrics, ReMAC surpasses its competitors. Notably, by learning missingness-invariant representations, our method maintains strong performance regardless of the percentage of missing values.

\begin{table}[ht]
\centering
\begin{adjustbox}{width=0.47\textwidth}
\begin{tabular}{cccccc}
\hline
\%MV                   & MODEL                                & ACC                                          & SEN                                 & SPE                                          & AUC                                          \\ \hline
                       & HistGB                               & 0.68 ± 0.15                                  & 0.50 ± 0.41                         & 0.83 ± 0.19                                  & 0.63 ± 0.16                                  \\
                       & XGB                                  & 0.68 ± 0.15                                  & 0.50 ± 0.41                         & 0.83 ± 0.19                                  & 0.58 ± 0.10                                  \\
                       & CatBoost                             & 0.72 ± 0.10                                  & \textbf{0.71 ± 0.34}                & 0.75 ± 0.17                                  & 0.76 ± 0.14                                  \\
\multirow{-4}{*}{0\%}  & \cellcolor[HTML]{EFEFEF}ReMAC (\textit{Ours}) & \cellcolor[HTML]{EFEFEF}\textbf{0.86 ± 0.10} & \cellcolor[HTML]{EFEFEF}0.67 ± 0.24 & \cellcolor[HTML]{EFEFEF}\textbf{1.00 ± 0.00} & \cellcolor[HTML]{EFEFEF}\textbf{0.86 ± 0.10} \\ \hline
                       & HistGB                               & 0.68 ± 0.28                                  & 0.63 ± 0.48                         & 0.75 ± 0.17                                  & 0.71 ± 0.28                                  \\
                       & XGB                                  & 0.68 ± 0.28                                  & 0.63 ± 0.48                         & 0.75 ± 0.17                                  & 0.69 ± 0.29                                  \\
                       & CatBoost                             & 0.77 ± 0.18                                  & \textbf{0.71 ± 0.34}                & 0.83 ± 0.19                                  & 0.81 ± 0.18                                  \\
\multirow{-4}{*}{10\%} & \cellcolor[HTML]{EFEFEF}ReMAC (\textit{Ours}) & \cellcolor[HTML]{EFEFEF}\textbf{0.82 ± 0.14} & \cellcolor[HTML]{EFEFEF}0.67 ± 0.24 & \cellcolor[HTML]{EFEFEF}\textbf{0.92 ± 0.17} & \cellcolor[HTML]{EFEFEF}\textbf{0.86 ± 0.10} \\ \hline
                       & HistGB                               & 0.68 ± 0.28                                  & 0.63 ± 0.48                         & 0.75 ± 0.32                                  & 0.69 ± 0.28                                  \\
                       & XGB                                  & 0.58 ± 0.17                                  & 0.50 ± 0.41                         & 0.67 ± 0.27                                  & 0.60 ± 0.18                                  \\
                       & CatBoost                             & 0.72 ± 0.10                                  & \textbf{0.71 ± 0.34}                & 0.75 ± 0.17                                  & 0.83 ± 0.19                                  \\
\multirow{-4}{*}{20\%} & \cellcolor[HTML]{EFEFEF}ReMAC (\textit{Ours}) & \cellcolor[HTML]{EFEFEF}\textbf{0.81 ± 0.16} & \cellcolor[HTML]{EFEFEF}0.67 ± 0.24 & \cellcolor[HTML]{EFEFEF}\textbf{0.92 ± 0.17} & \cellcolor[HTML]{EFEFEF}\textbf{0.86 ± 0.10} \\ \hline
                       & HistGB                               & 0.73 ± 0.22                                  & \textbf{0.71 ± 0.34}                & 0.75 ± 0.17                                  & 0.67 ± 0.24                                  \\
                       & XGB                                  & 0.68 ± 0.15                                  & \textbf{0.71 ± 0.34}                & 0.67 ± 0.00                                  & 0.67 ± 0.19                                  \\
                       & CatBoost                             & 0.72 ± 0.10                                  & 0.58 ± 0.29                         & 0.83 ± 0.19                                  & \textbf{0.89 ± 0.08}                         \\
\multirow{-4}{*}{30\%} & \cellcolor[HTML]{EFEFEF}ReMAC (\textit{Ours}) & \cellcolor[HTML]{EFEFEF}\textbf{0.81 ± 0.16} & \cellcolor[HTML]{EFEFEF}0.54 ± 0.42 & \cellcolor[HTML]{EFEFEF}\textbf{1.00 ± 0.00} & \cellcolor[HTML]{EFEFEF}0.86 ± 0.10 \\ \hline
\end{tabular}
\end{adjustbox}
\caption{\textit{Performance of classification models capable of handling missing values, together with ReMAC, under different missing-value (\%MV) regimes. Metrics include accuracy (ACC), sensitivity (SEN), specificity (SPE), and area under the curve (AUC). The best value for each metric is highlighted in bold, and the proposed method is shown with a gray background.}}
\label{tab:results_incomplete_data}
\end{table}

\subsection{Analysis of representation strategies}

\begin{figure}[h]
\centering
        \includegraphics[width=0.45\textwidth]{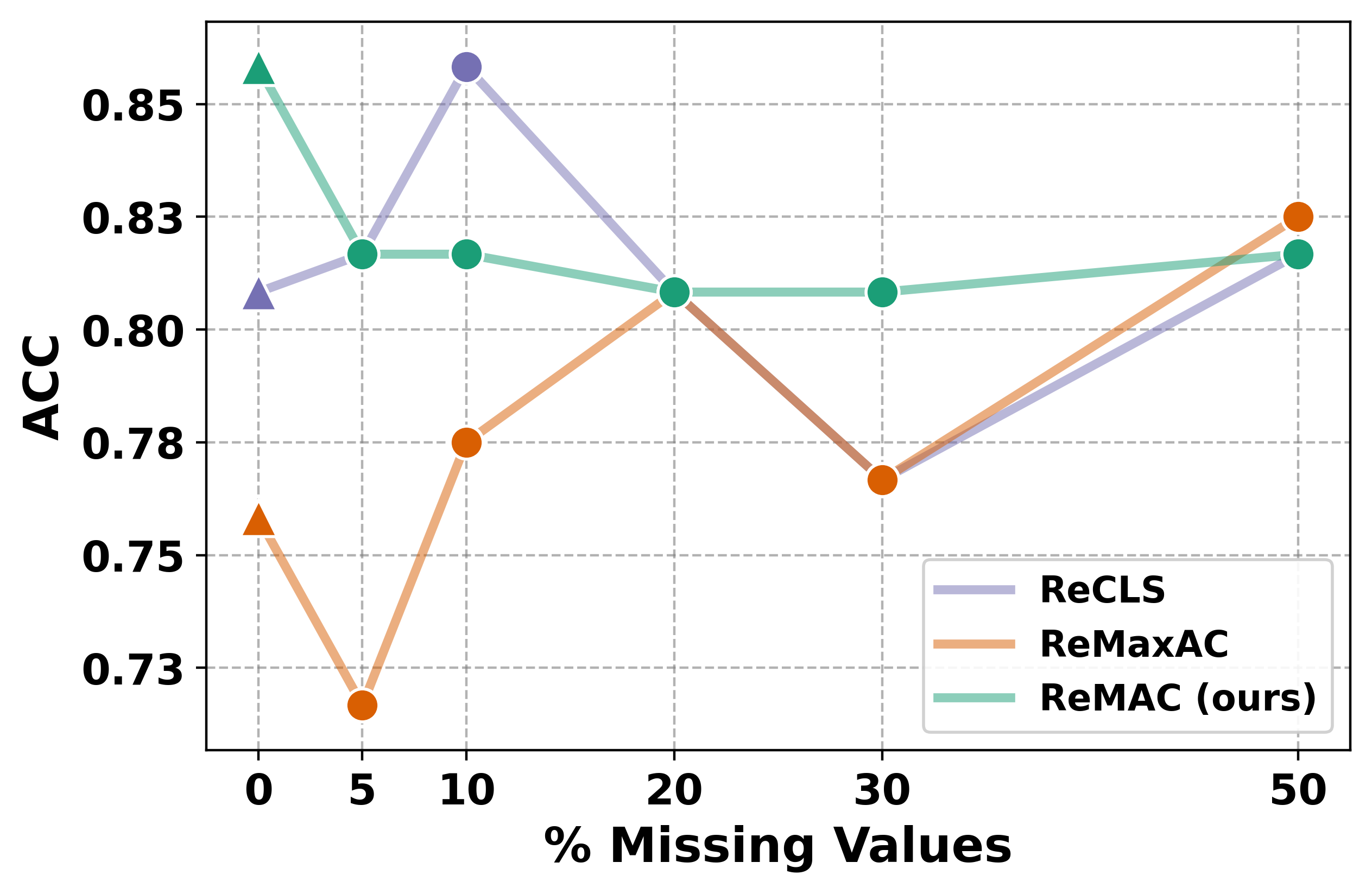}
    \caption{\textit{Comparison of different representation strategies for decoder embeddings under varying missing-value (\%MV) regimes. Accuracy (ACC) is reported for ReCLS (learnable token), ReMaxAC (max pooling), and ReMAC (mean pooling, ours).}}
    \label{fig:aggs}
\end{figure}

Different strategies for aggregating the latent representations of the decoder were explored for the classification task. Specifically, we considered the use of a learnable embedding appended to the sequence of Transformer decoder blocks (ReCLS), similar to the approach commonly adopted in Vision Transformers, as well as aggregation of decoder embeddings through max pooling (ReMaxAC) and mean pooling (ReMAC). The results are depicted in Fig. \ref{fig:aggs}. It is worth noting that ReMAC not only achieves the best performance in the absence of missing values but also exhibits the highest stability across different missing-value regimes. Consequently, in this high-dimensional setting, mean pooling yields a more robust and discriminative latent representation space compared to other, seemingly more straightforward, alternatives such as ReCLS.

\subsection{Impact of mask ratio and classification architecture}

\begin{figure}[ht]
     \centering
     \begin{subfigure}[b]{0.45\textwidth}
         \centering
         \includegraphics[width=\textwidth]{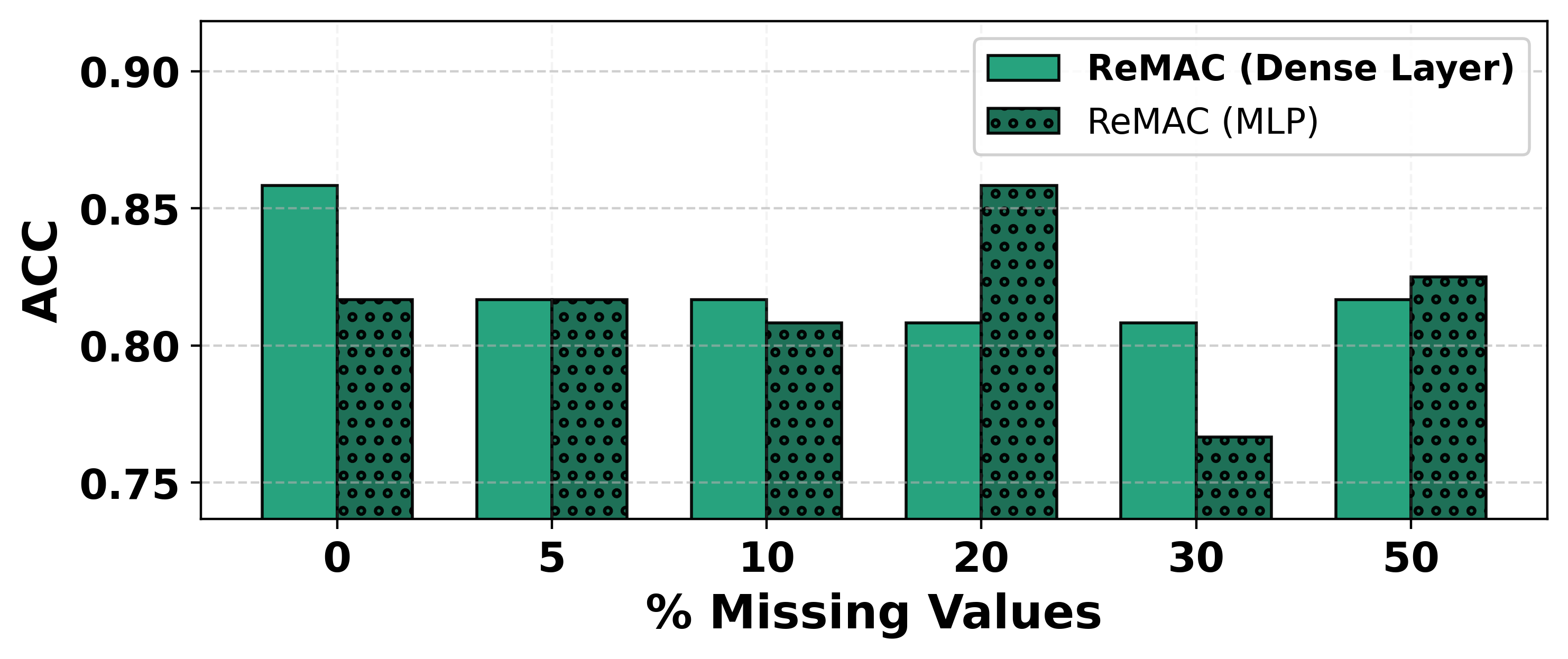}
         \caption{\centering\textit{Effect of classification head design: dense layer vs. MLP.}}
         \label{fig:clf_comparison}
     \end{subfigure}
     \hfill
     \begin{subfigure}[b]{0.45\textwidth}
         \centering
         \includegraphics[width=\textwidth]{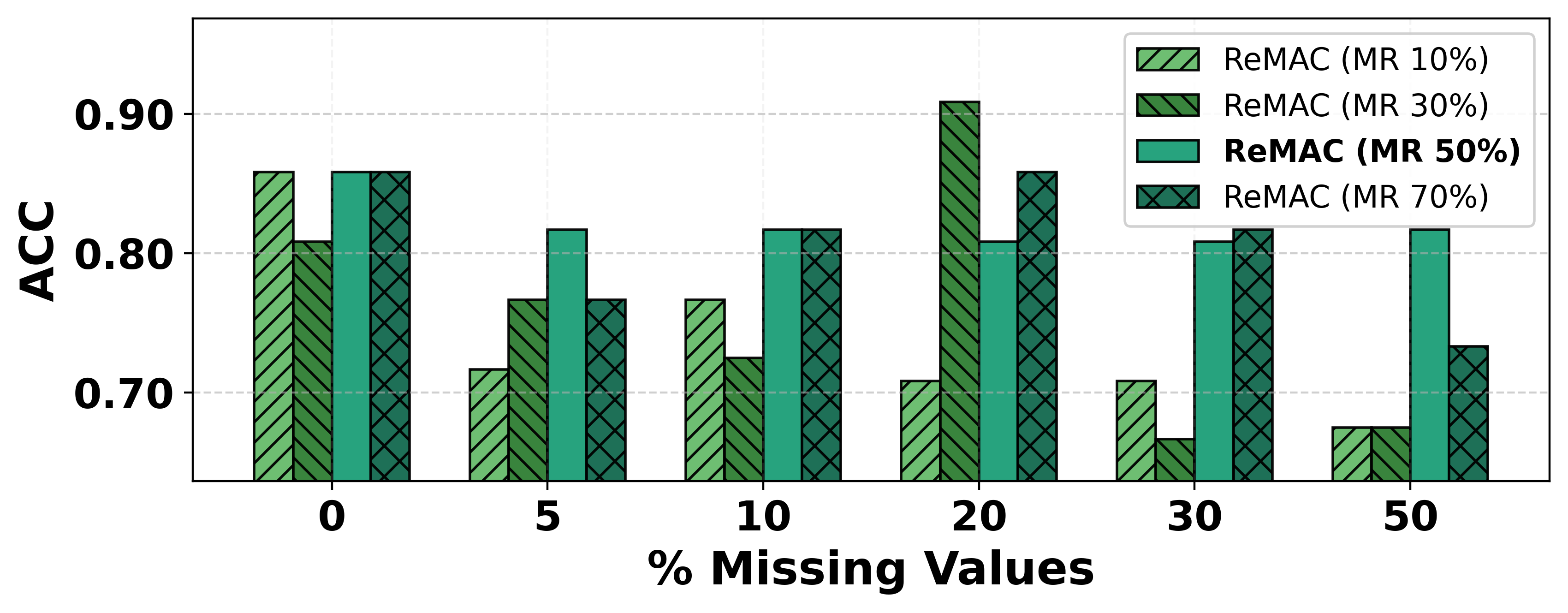}
         \caption{\centering\textit{Effect of different mask ratios (MR).}}
         \label{fig:mr_comparison}
     \end{subfigure}
        \caption{\textit{Ablation Studies on ReMAC: impact of the classification head complexity and the mask ratio (MR) under varying missing-value regimes. In both subfigures, the configuration adopted in the main experiments is highlighted in bold in the legend.}}
        \label{fig:ablation_studies}
\end{figure}

Figure \ref{fig:ablation_studies} presents the results of the ablation studies evaluating the architectural design of the classification head and the mask ratio (MR) used to mask input features randomly. Regarding Fig. \ref{fig:clf_comparison}, results indicate slightly better and more consistent performance when employing a single dense layer compared to a multilayer perceptron (MLP) with one hidden layer. This shows that a simple classification head is sufficient, providing stable generalization across multiple scenarios due to an already discriminative latent space. Concerning Fig. \ref{fig:mr_comparison}, we observe that, across different missing-value regimes, ReMAC generally performs better with higher mask ratios (MR). These results are align with the intuition that larger mask ratios, by exposing the model to a greater number of masked variables, naturally encourage more consistent representation learning under missing values.

\vspace{-1mm}
\section{Conclusion}

The diagnosis of spitzoid tumors is essential for ensuring timely and appropriate treatment. Epigenetic data, such as DNA methylation, have emerged as a valuable source of information for ST stratification. However, the frequent presence of missing values in data collection highlights the need for effective methodologies under such regimes. In this work, we introduced ReMAC, an extension of ReMasker tailored for high-dimensional classification tasks. Our results demonstrate that the proposed method achieves strong performance both on complete data and under varying degrees of incompleteness. The main limitations of this study include the small sample size, which is inherent to the problem under investigation, and the need to further assess the explainability of the method. These findings open a promising avenue for extending ReMAC to larger datasets and exploring its potential in providing interpretable predictions.

\section*{Acknowledgments}

This work has received funding from the Spanish Ministry of Economy and Competitiveness through project PID2022-140189OB-C21 (ASSIST) and has also been supported by the Generalitat Valenciana via projects INNEST/2021/321 (SAMUEL), CIPROM/2022/20 (PROMETEO), and SA2024-26 (IMAGEN).

\begin{small}
\bibliographystyle{ieeetr}
\bibliography{references}

@article{remasker,
  title={Remasker: Imputing tabular data with masked autoencoding},
  author={Du, Tianyu and Melis, Luca and Wang, Ting},
  journal={arXiv preprint arXiv:2309.13793},
  year={2023}
}

@inproceedings{delamor2021deep,
  title={A deep embedded framework for spitzoid neoplasm classification using dna methylation data},
  author={Del Amor, Roc{\'\i}o and Colomer, Adri{\'a}n and Monteagudo, Carlos and Garz{\'o}n, Mar{\'\i}a Jos{\'e} and Garc{\'\i}a-Gim{\'e}nez, Jos{\'e} Luis and Naranjo, Valery},
  booktitle={2021 29th European Signal Processing Conference (EUSIPCO)},
  pages={1271--1275},
  year={2021},
  organization={IEEE}
}

@article{gonzalez2023diagnostic,
  title={Diagnostic Algorithm to Subclassify Atypical Spitzoid Tumors in Low and High Risk According to Their Methylation Status},
  author={Gonz{\'a}lez-Mu{\~n}oz, Jose Francisco and S{\'a}nchez-Sendra, Beatriz and Monteagudo, Carlos},
  journal={International Journal of Molecular Sciences},
  volume={25},
  number={1},
  pages={318},
  year={2023},
  publisher={MDPI}
}

@article{barnhill2006spitzoid,
  title={The Spitzoid lesion: rethinking Spitz tumors, atypical variants,‘Spitzoid melanoma'and risk assessment},
  author={Barnhill, Raymond L},
  journal={Modern pathology},
  volume={19},
  pages={S21--S33},
  year={2006},
  publisher={Elsevier}
}

@article{orchard1997spitz,
  title={Spitz naevi misdiagnosed histologically as melanoma: prevalence and clinical profile},
  author={Orchard, David C and Dowling, John P and Kelly, John W},
  journal={Australasian journal of dermatology},
  volume={38},
  number={1},
  pages={12--14},
  year={1997},
  publisher={Wiley Online Library}
}

@article{van2011mice,
  title={mice: Multivariate imputation by chained equations in R},
  author={Van Buuren, Stef and Groothuis-Oudshoorn, Karin},
  journal={Journal of statistical software},
  volume={45},
  pages={1--67},
  year={2011}
}

@article{kyono2021miracle,
  title={Miracle: Causally-aware imputation via learning missing data mechanisms},
  author={Kyono, Trent and Zhang, Yao and Bellot, Alexis and van der Schaar, Mihaela},
  journal={Advances in Neural Information Processing Systems},
  volume={34},
  pages={23806--23817},
  year={2021}
}

@article{hivae,
  title={Handling incomplete heterogeneous data using vaes},
  author={Nazabal, Alfredo and Olmos, Pablo M and Ghahramani, Zoubin and Valera, Isabel},
  journal={Pattern Recognition},
  volume={107},
  pages={107501},
  year={2020},
  publisher={Elsevier}
}

@inproceedings{yoon2018gain,
  title={Gain: Missing data imputation using generative adversarial nets},
  author={Yoon, Jinsung and Jordon, James and Schaar, Mihaela},
  booktitle={International conference on machine learning},
  pages={5689--5698},
  year={2018},
  organization={PMLR}
}

@inproceedings{arik2021tabnet,
  title={Tabnet: Attentive interpretable tabular learning},
  author={Arik, Sercan {\"O} and Pfister, Tomas},
  booktitle={Proceedings of the AAAI conference on artificial intelligence},
  volume={35},
  number={8},
  pages={6679--6687},
  year={2021}
}

@article{beck2022genome,
  title={Genome-wide CpG density and DNA methylation analysis method (MeDIP, RRBS, and WGBS) comparisons},
  author={Beck, Daniel and Ben Maamar, Millissia and Skinner, Michael K},
  journal={Epigenetics},
  volume={17},
  number={5},
  pages={518--530},
  year={2022},
  publisher={Taylor \& Francis}
}

@article{seiler2021characterizing,
  title={Characterizing the properties of bisulfite sequencing data: maximizing power and sensitivity to identify between-group differences in DNA methylation},
  author={Seiler Vellame, Dorothea and Castanho, Isabel and Dahir, Aisha and Mill, Jonathan and Hannon, Eilis},
  journal={BMC genomics},
  volume={22},
  number={1},
  pages={446},
  year={2021},
  publisher={Springer}
}
\end{small}

\end{document}